\documentclass[12pt]{article}
\usepackage{epsf}
\begin{document}
\baselineskip 8 mm

\begin{center}
\Large{\bf Nucleus-nucleus collisions
in the Dynamical String Model}\\[50pt]

\normalsize
  $^{a,b}$B. Iv\'anyi, $^a$Z. Schram, $^a$K. Sailer 
   and $^b$G. Soff\\[25pt]

 $^a$Department for  Theoretical Physics,\\
   Kossuth Lajos University,\\
  H-4010 Debrecen, Pf. 5, Hungary\\[15pt]
 $^b$Institute for Theoretical Physics,\\
   Technical University,\\
  D-01062 Dresden, Germany\\

\end{center}

\begin{abstract}

In this paper the Dynamical String Model is applied to the numerical
simulation of the ultrarelativistic heavy-ion collision 
$^{32}$S(200 GeV/n)$+^{32}$S. The results 
are in qualitative agreement with experimental data.

\end{abstract}

\section{Introduction}

The purpose of the present paper is to show that the Dynamical String
Model  offers an alternative approach to describe ultrarelativistic heavy-ion
collisions at bombarding energies of a few hundreds GeV/nucleon
 in terms of extended objects, the so-called hadronic
strings.

The existing event generators for high-energy hadronization processes
can be classified as follows:
\begin{enumerate}
\item Event generators for high-precision description of elementary 
hadronization processes (including only lepton and proton beams) in vacuum:
PYTHIA \cite{Pythia}, HERWIG \cite{Herwig}, ARIADNE \cite{Ariadne},
LEPTO \cite{Lepto}, ISAJET \cite{Isajet}.
These event generators combine the parton-shower evolution in
perturbative QCD terms with non-perturbative hadronization
prescription to convert final partonic distributions into hadronic ones.
For the latter the LUND string fragmentation model is commonly used
 \cite{And83}
with the exception of HERWIG, which  uses other  considerations for coalescing
coloured partons to colour-neutral clusters and fragmenting those into
 hadrons. The common feature of this
class of models is the lack of space-time evolution. Therefore, these
models cannot be directly applied for describing hadronization at
finite densities, e.g.\ for that of high-energy hadronization
processes involving also nuclei.
\item Event generators for the description of hadronization at finite
densities, e.g.\ high-energy heavy-ion collisions: FRITIOF
\cite{Fritiof}, PCM \cite{Pcm}, DPM \cite{Dpm}, VENUS \cite{Venus},
QGSM \cite{Qgsm}, RQMD \cite{Rqmd}, UrQMD \cite{Urqmd}, HSD \cite{Hsd}, 
HIJET \cite{Hijet}, HIJING \cite{Hijing},
 Two-phase simulation of ultrarelativistic nuclear 
collisions \cite{Kahana}.

These models provide a space-time description (except HIJING)
of the hadronization process considered. In a few of them
\cite{Pcm,Dpm,Kahana, Hijing}  the
partonic degrees of freedom are included also in the early stage of the 
collision in some form. Hadrons are generally treated as point
particles with interaction ranges prescribed
on the base of the constituent valence-quark picture. As a rule,
string excitation  is included (with the exception of
\cite{Kahana}) in various forms and the LUND string fragmentation
model \cite{And83} is used. Generally, string-like excited hadronic
states do not propagate and collide with other hadrons in the
surrounding. Strings are rather a clever tool of bookkeeping how
highly excited hadrons fragment into hadrons of the discrete mass spectrum.

\item Models which intend to provide a space-time description for both the
 high-energy elementary hadronization processes in vacuum and the more involved
 hadronization processes at finite densities like those including
 nuclei, e.g.\ VNI \cite{Vni}.
 VNI gives a full space-time picture of ultrarelativistic heavy-ion
 collisions, combining the space-time evolution of the parton-shower
 in its early stage with the later hadronic cascade. The space-time
 picture of the parton-cascade has also been used in the parton-hadron
 conversion, based on the ideas introduced in \cite{Herwig}. There are
 no strings in this model, 
the hadrons are considered as  point particles with finite interaction
ranges, as usual.
\end{enumerate}

The Dynamical String Model presented here belongs to the third class
of models providing a full space-time description. It can be applied
to elementary hadronization processes in vacuum and to hadronization
processes at finite densities
involving nuclei as well.
 It is the basic feature of the Dynamical String Model that
during the whole space-time evolution of an event all the hadrons are
consistently considered as extended, string like  objects satisfying 
the  particular laws of string dynamics. On the
 contrary to other models on the market, 
  the string picture is not merely used as a
 fragmentation model of excited hadrons. It is taken here as the model
 for hadron dynamics,
according to which the laws of motion,  decay, and 
collision of hadrons are determined.  

On the contrary to the models including a parton-shower
for the early stage of the evolution, the Dynamical String Model considers
the string like collective excitations of the hadrons to be decisive for the
evolution of the ultrarelativistic heavy-ion collision, neglecting
completely the underlying partonic processes. The good  qualitative
 description of ultrarelativistic heavy-ion
collisions for CERN SPS energies of a few hundreds GeV/n obtained 
in the present paper  indicates that the overall
qualitative features of the fragment distributions and multiplicities
may not be sufficient to clarify the interplay of the string like
collective degrees of freedom and that of the partonic ones. 

In the Dynamical String Model all kinds of broken line string
excitations are taken into account, whereas in other existing models
string like excitations are basically longitudinal, yo-yo like
as far as no gluon jets (or minijets) are included.

The Dynamical String Model has a rather few number of parameters, as
compared to other existing models. That is an advantage, but on the
other hand one cannot expect that the model in this form can
provide more than an overall qualitative description of
ultrarelativistic heavy-ion collisions.

In Sect.\ 2 we give a description of the Dynamical String Model, and in
Sect.\ 3 the model is applied to the ultrarelativistic heavy-ion 
collision $^{32}$S(200 GeV/n)+$^{32}$S.

\section{Dynamical String Model}

\subsection{Motivation}

The underlying idea of the Dynamical String Model \cite{kornelrew}
is  that hadrons can be represented by classical one-dimensional objects,
 the oriented
relativistic open bosonic strings as suggested by Artru \cite{Art83}
 and Remler \cite{Remler}.  
There are experimental evidences that hadrons  have string like
collective degrees of freedom: (i) the well-known, almost linear
 Regge-trajectories \cite{Coll71} corresponding to the string tension
of $\kappa \approx 0.9$ GeV/fm, (ii) the nearly exponential mass
 spectrum of the resolved hadron resonances \cite{Hag65}; (iii) the
 existence of a preferred (longitudinal) direction in elementary
 fragmentation processes; (iv) the emission of linearly polarized
 gluons by the excited hadronic system occurring in high-energy pp collisions
 \cite{Jac90}. Theoretical indications and successful applications
of the string model for hadronic physics are overviewed in
\cite{Sai91}. For our work the success of the
string fragmentation models developed by Artru and Mennessier
\cite{Art74} and by the Lund group \cite{And83} was particularly
encouraging.
The oriented relativistic open string is thought of the idealization
of the chromoelectric flux-tube with quark and antiquark (diquark)
ends for mesons (baryons). The endpoints are assumed to have vanishing
rest masses.
The original idea is then modified: the infinitesimally thin
strings have been replaced with more realistic thick ones, i.e.\ 
with strings
exhibiting a finite tranverse size, more precisely a radius $R$. 
The hadronic strings introduced in this manner are treated 
afterwards in a fully dynamical
way in our model. They propagate, collide and decay according to the
particular laws deduced from the string picture and from the analogy
of hadronic strings with chromoelectric flux tubes, as described below.

Energy and momentum conservations are strictly satisfied in any
elementary decay and collision event and in the evolution of the whole
hadronic system, as well. No spin is introduced and the angular
momentum conservation is not considered.

\subsection{Mass spectrum}

Our starting point is the classical Nambu-Goto string \cite{Nam70,hatfield}.
The classical mechanical string picture provides us with a continuous mass
spectrum. All kinds of broken line string configurations
can arise during the evolution of any system of hadronic strings due
to  inelastic string collisions. Furthermore, it has also been shown
that  those broken line string
configurations with arbitrary number of kinks
are unavoidable to obtain a realistic exponential mass spectrum of
hadronic strings \cite{Saimul89}. A finite amount
 of momentum (and energy) can be carried by the kinks and 
by the string endpoints as well.

 In order to be more realistic, below the mass
thresholds  of 1.5 GeV for baryons and 1.0 GeV  for mesons
strings only with discrete rest masses taken from \cite{PDG}
are allowed in the model. Strings in the rotating rod mode
are associated to the discrete hadronic states which correspond 
to the leading Regge-trajectories \cite{Art83}. 
 It holds $2M= \kappa \pi \ell$ for their
lengths $\ell$ and rest masses $M$ leading e.g. to $\ell \sim $0.7 fm for the
nucleon, and  $\ell \sim $0.1 fm for the pion. Particles containing
strange valence quarks are completely neglected.

The string endpoints carry the appropriate baryonic charges and baryon
number conservation is satisfied. 
Spin of the hadronic strings, electric charges and flavours of the string
endpoints have not been introduced. On the other hand, the degeneracies
of the discrete resonance states due to their spin and isospin are
taken into account.

\subsection{Free motion}

Any string configuration can be encoded in the trajectory of one of
the endpoints of the string, in the so-called directrix \cite{Art83},
and boosted to any requested velocity as described in \cite{kornelrew}
in detail. The directrix determines the string configuration at a given 
time and also its free evolution according to the Nambu-Goto action.
Any influence of the assumed transverse size of the hadronic strings
on their free motion is neglected. During the evolution of the
investigated system
each hadronic string is assumed to move freely between the subsequent
elementary interaction events (decays and collisions) of its life.  

In the numerical code the directrix is stored in less than 200 points,
with typically 0.1 GeV rest mass for every linear segment of the
broken line string. Whenever it is needed for describing single
decay and collision events, the string can be
constructed from its directrix unambigously \cite{Art83,kornelrew}.
The string endpoints generally carry a finite amount of momentum,
and are described by two string points at the same spatial position,
but corresponding to different values of the string parameter
\cite{kornelrew}.

In order to simulate any individual elementary string interaction
event, the participating strings are reconstructed numerically from
their directrices.
After carrying out a single decay or collision event the final state
strings must be converted back in their directrices. Generally, the
conversion directrix $\rightarrow$ string $\rightarrow$ directrix
leads to a doubling of the directrix points with many
redundant ones  that have to be removed by a reduction algorithm.
If the endpoints of two neighbouring directrix segments are almost on a
straight line (i.e.\ the common point of both segments is rather close to
one of the endpoints, or to the straight line connecting the endpoints),
the segments are replaced by a single linear directrix segment
under the constraint that energy and momentum must be conserved.

\subsection{String collision}

 In the Dynamical String Model the collision is introduced as a binary
interaction of strings.
In order to obtain realistic total cross sections for the
string-string collisions, a finite transverse size or radius $R$ has to
be prescribed to the hadronic strings as already established in
\cite{kornelrew}. This radius is chosen
to be identical for all hadronic strings and it is  also assumed not to be
Lorentz contracted \cite{Bjo76}. Strings coming in touch, and 
remaining after a critical collision time $\tau_c$ still closer than their
interaction range $2R$, interact.
The total cross section is assumed to
have a purely geometrical origin. Elastic and inelastic string-string 
collisions are distinguished, based also on geometrical
concepts. An inelastic interaction range $R'<R$ is defined. Strings
being closer than $2R'$ after the time $\tau_c$ elapsed since they came
in touch,
 suffer inelastic collision, whereas the
peripheral collisions are considered to be elastic ones.
If the strings came in touch but after the time $\tau_c$ they are
at a distance larger than $2R$, they do not interact.  

\begin{figure}
\vskip -1.5cm
\epsfxsize=1.0\columnwidth
\epsffile{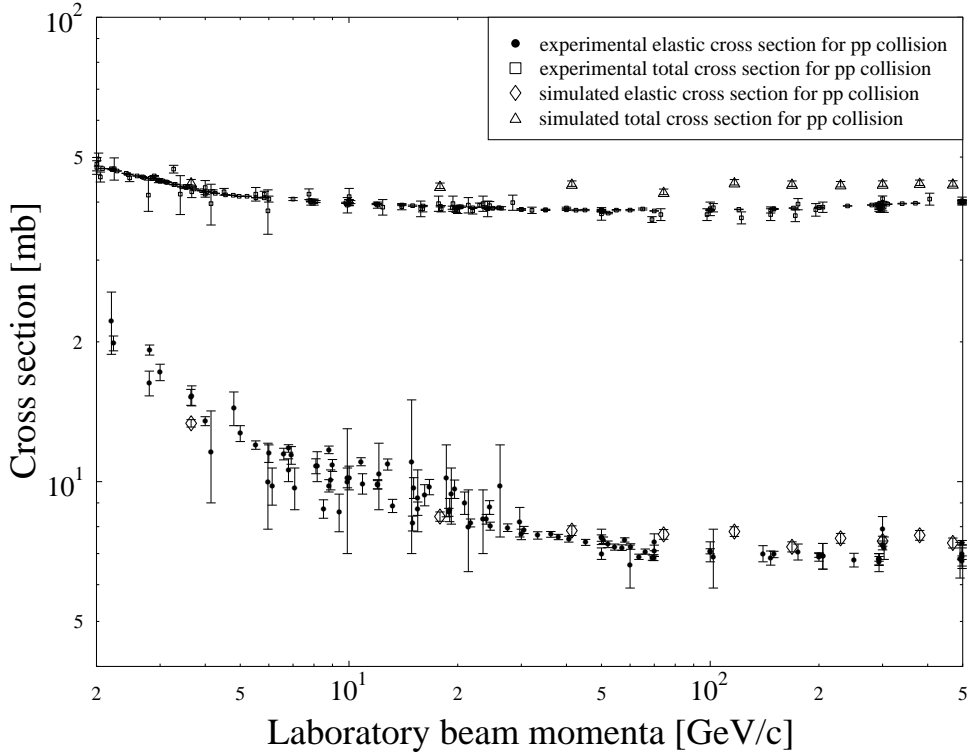}
\caption{\label{sigma}Simulated total and elastic cross section for $pp$
collision at different beam momenta.}
\end{figure}

 The differentiation between elastic and inelastic processes
described above was tested by numerical simulation of proton-proton
($pp$) collisions  in the energy range
$\sqrt{s}=3-30$ GeV (see Fig.\ \ref{sigma}).
For determining the total and elastic $pp$
cross sections $10^4$ collision events were numerically simulated
by shooting a projectile proton ($N_p =1$) on a target proton ($N_t
=1$)  at rest with an
impact parameter $\le \rho$. That is to say, the center of the 
target proton was
positioned on the beam axis, the centers of the projectiles were unifomly
distributed on a disk of radius $\rho \approx 1$ fm centered on the
beam axis in the transverse plane. The initial states of the protons were
represented by rotating rod modes.
The orientations of the projectile
and the target protons were uniformly distributed in the entire solid
angle. Projectile protons were produced at a longitudinal distance 
3 fm from the target. The simulations were performed by using the time
steps $\Delta t = 0.02$ fm/c. The numbers $N_{tot}$ and $N_{el}$ of
events  with interaction and with elastic collision, respectively, were
counted and converted to the corresponding
total and elastic cross sections, $\sigma_{tot} = \rho^2 \pi N_{tot} /
 (N_t N_pv_p)$ and  $\sigma_{el} = a^2 N_{el} / (N_t N_p v_p)$   
with the projectile velocity $v_p$.

It is more involved (and also more time consuming) to calculate
 the distance of two strings than to determine the
distance  of two point particles. The  distance $d$ 
of two colliding strings is defined as
the minimal distance of the points of a projectile
string $a$ and those
of the target string $b$. The distance $d$ is monitored for each
string pair $a$ and $b$  in every time step as follows. 

The  distance between strings $a$ and $b$ can be
estimated as $d_{\rm est} = |\vec{x}_a - \vec{x}_b| - \frac{1}{2}(l_a + l_b)$,
where $\vec{x}_{a,b}$ are the centers of mass and $l_{a,b}$ 
are the lengths of the
strings $a$ and $b$, respectively. If $d_{\rm est}$ is larger than the string
interaction range
$2R$, then $d_{\rm est}$ is taken for the distance of  the string pair.
 Otherwise, the real
 distance $d$ is computed. If once the real distance $d$ is calculated
for a string pair and turned out to be larger than the interaction
range, $d > 2R$, it is not
 checked up to the time $t = \frac{d - 2R}{2c}$
with  $c$ the speed of light.

After the strings came in
touch, i.e.\ their distance became $d (t_0) \le 2R$ at time $t_0$, they can
interact. The decisions on the interaction and on the
interaction  channel (if any) are taken after a critical time $\tau_c$ elapsed.
There is no interaction if the distance of the strings 
$d (t_0 + \tau_c) >2R$, while  an elastic collision or an inelastic collision
takes place if  $2R > d (t_0 + \tau_c) > 2R'$ or 
$d (t_0 + \tau_c) < 2R'$, respectively.

 It was concluded that the
elastic and inelastic
cross sections shown in Fig.\ \ref{sigma}
 are in qualitative agreement with  experimental data
for the interaction ranges $R\approx 0.6$ fm and $R'=0.7\,R \approx
0.4$ fm and the interaction time $\tau_c = 0.4$ fm/c. The ratio
 of the simulated elastic and total cross sections, and the energy
 dependence of the elastic cross section are  rather
sensitive to the ratio $R'/R$ and to $\tau_c$.
The simulated results are consistent with  the following order of
magnitude estimates valid
for large values of the projectile momentum:
 $\sigma_{tot} \rightarrow
4R^2 \pi \approx 45$ mb and $\sigma_{el} \rightarrow 4(R-R')^2 \pi \approx
5$ mb. These  estimates do not take into account the orientations
of the strings.
There is a difference of the results obtained here
($\sigma_{tot}= 45$ mb for $R=0.6$ fm) and in \cite{kornelrew}
(the same $\sigma_{tot}$ for $R=0.45$ fm). It is the consequence of
introducing the interaction time $\tau_c$. Strings overlapping only at
their ends for a short time may leave their interaction range during
the time $\tau_c$ after having come in touch.

In the Dynamical String Model the distance of any pair of strings must
be determined in every time step with the algorithm described above.
A great amount of computational time can be spared in the simulation
of heavy-ion collisions by determining the
estimate $d_{\rm est}$ and not calculating the actual distance for
$d_{\rm est} >2R$.
Once in the simulation of a heavy-ion collision event
 a string pair came in touch and the time $\tau_c$ is over,
 the channel of the collision (inelastic,
elastic, or no interaction) is decided as described above and the
appropriate final state in the actual collision channel is generated.

In the simulations presented here the elastic channel is introduced
only to reduce the inelastic fraction of the total cross section, but
the strings that suffered an elastic collision were let to move
further as if nothing had happened.

The inelastic collisions are considered as rearrangements
\cite{Art83}. The rearrangement of infinitely thin colliding strings 
is a simple cut followed by the reconnection  of the string arms at 
the point of  intersection.  The order of the reconnection is always unique,
as the strings are oriented objects. In the numerical code the rearrangement
is carried out in the time step when the interaction time $\tau_c$
after the strings came in touch is over, and the criterium $d < 2R'$ is
fulfilled.  Then the points of the minimal distance define the points
 where the strings are disjoined and reconnected once again.
The reconnection is performed by displacing the appropriate string
pieces. Energy and momentum are conserved automatically, and the center of
energy of the string pair is conserved by displacing the string pair
as a whole appropriately. Owing to the  interaction time $\tau_c$ the
new strings generally can move away, so that any  undesired infinite
 sequence of interactions is avoided.

According to the above prescription of rearrangement for string with
continuous mass spectrum,
it may happen that one or both of the final state strings would have
rest masses below the mass thresholds. Similar final states could
also arise as the result of decay. Their treatment shall be discussed
later in detail. 

Collision of discrete resonances or that of a discrete resonance with
a string of the continuous mass spectrum are treated according to
the same rules as the collisions of the strings of the continuous mass
spectrum, as the resonances are represented by strings in the rotating
rod mode.

\subsection{Decay}

The decay law for  relativistic strings belonging to the continuous
mass spectrum is given by $dw = -\Lambda dA$, i.e. the probability
$dw$ that the string piece  having swept the invariant area $dA$
breaks is proportional to that area, with the decay constant $\Lambda$
\cite{Art74}. Making use of the analogy of the hadronic string with
the chromoelectric flux tube, the decay is considered as the result
of the production of a quark-antiquark pair via  tunneling effect
 in the strong
chromoelectric field of the flux tube \cite{Cas79}. Then the decay constant
$\Lambda = R^2 \pi w (R)$ can 
be expressed in
terms of the quark-antiquark pair production rate $w(R)$ depending on
the radius of the flux tube \cite{kornel1,Pav91}. The created quark
and antiquark acquire oppositely directed  transverse momenta with
an approximately Gaussian  distribution  $ dP (p_T) \sim \exp ( - p_T^2/
2p_{T0}^2) dp_T^2$ $(p_{T0} \approx 1.43/R)$ deduced from the analogy with
flux tubes \cite{Schon90}. 

The decay of strings with rest masses above the threshold are
simulated as follows. For any string created, an invariant area $A_0$
is chosen according to the distribution $\sim \exp ( - \Lambda A_0)$.
The increment of the invariant area swept by the string is calculated
in every time step as the sum of area elements of the linear string
segments. In the time step when the invariant area swept by the string
exceeds the value $A_0$, the string is broken up without any time delay.
 The decay is performed in the segment for which
the probability of decay has a maximum for that time step.
The transverse momenta of the new string ends are chosen with the
distribution $dP(p_T)$, and with uniform distribution in the 
plane transverse to the decaying string piece in its rest frame.
 A piece of the string is removed around the breaking point
that is required to satisfy energy and momentum conservation.

The decay of discrete resonances is not considered like a string
decay. Their lifetimes and decay channels  are taken
from \cite{PDG}. According to the exponential decay law of point
particles a time $T_0$ is chosen for every resonance created, and
having that elapsed its decay is performed.

The decay of discrete  resonances can result in two or three
daughters. For decay into two daughters the magnitudes of their
momenta are well-defined, and the direction of the momenta is chosen
isotropically in the rest frame of the mother resonance.
For decay into three daughters it is assumed that the momenta of the
daughters lie in a plane of randomly chosen orientation in the rest
frame of the mother resonance, the momenta are of equal magnitude 
and each neighbouring
 pair closes the angle
 120$^{o}$ in the rest frame of the mother. The direction of one
 of  the three momenta are chosen
 isotropically in the plane. The magnitudes of the momenta determined
under these assumptions from energy conservation 
are in agreement with the average momenta indicated in \cite{PDG}.
Finally, the daughter resonances are represented by the appropriate
rotating rods, displaced out of their interaction range in the
directions of their momenta.

\subsection{Final states with discrete resonances}

Both rearrangement  and decay of strings belonging to the
continuous mass spectrum  can result in two-particle
final states
with one or both  strings below the mass threshold.
 These cases are treated in the
model in different ways. 
\begin{enumerate}
\item The case with two rest masses below the threshold
is considered  as a final state with two discrete resonances.
The pair of resonances is chosen randomly according to the degeneracies 
of the resonances, under the restriction that the sum of the rest
masses of the resonance pair must not exceed the invariant mass of the
colliding string pair (of the mother string). Then the momenta of 
the resonances are chosen
randomly with isotropic orientation in the rest frame of the pair,
satisfying energy and momentum conservation. Finally, the resonances
are represented by rotating rods with the appropriate rest masses and
momenta and displaced in the direction of their momenta out of their
interaction range conserving the center of energy of the pair.
\item  The case with one rest mass $m_r$ below the threshold
is considered  as a final state with one discrete resonance
and a string belonging to the continuous part of the mass spectrum.
The resonance is chosen randomly taking the degeneracies into account,
under the restriction that the sum of the rest masses of the final
state particles must not exceed the invariant mass of the initial
strings (of the mother string). Further on, one has to proceed
differently for rearrangement and for decay.

\begin{enumerate}
\item For rearrangement:
One has to distinguish the cases with a mesonic or baryonic string
occurring below the corresponding mass threshold.
\begin{enumerate}
\item Meson below the mass threshold.
If the rest mass $m_r$ below the mesonic threshold $M_M$ turned out to be
smaller
 than the
pion mass, $m_r < m_\pi$, the colliding strings are considered to fuse
in  a single one. If $m_\pi < m_r < M_M$, the resonance with rest mass
$M_r < m_r$ but closest to $m_r$ is chosen with the same momentum
${\vec P}_r$ what the string with rest mass $m_r$ would have had.
The other string is slightly modified by chopping off its wedge at the
point of reconnection and inserting a linear segment of vanishing
momentum with rest mass $m_r- M_r$. 
\item Baryon below the mass threshold. Then the possibility of the
  fusion of both colliding strings is excluded in order to avoid 
exotic many quark states. Therefore, even if $m_r$ is smaller than the
nucleon mass, $m_r <m_N$, the proton is chosen for the discrete state.
The construction of the final state is performed similarly to that for
a discrete meson and a string. The mass difference $m_N - m_r$,
however, is now  taken away from the other string by chopping off its
wedge and displacing its arms to bring them in connection at their
 new endpoints. 
 Otherwise, for $m_N < m_r < M_B$ (with the baryonic threshold $M_B$) 
the final state is constructed in the same way as for a discrete meson
state and a string. 
\end{enumerate}

\item  For decay:
The smallest possible piece at the end of the continuous string
 is chopped off that is required to satisfy energy and momentum
conservation for the final state, when the resonance and the new
string endpoint acquire  the
 transverse momenta ${\vec p}_T$ and  $-{\vec p}_T$, resp.
The transverse momenta are
chosen randomly according to the distribution $dP(p_T)$, and oriented
isotropically in the plane perpendicular to the string at its endpoint.  
\end{enumerate}

Finally, the discrete resonance  is represented by the corresponding
 rotating rod and positioned so that the centers of energy of the
 initial and final states must be identical.
\end{enumerate}

\subsection{Parameters}

There are relatively few parameters in our model.
The Dynamical String Model has two basic parameters: the string
tension $\kappa \approx 0.9$ GeV/fm fitted to the slopes of the
leading Regge-trajectories, and the string radius $R \approx 0.6$ fm,
fitted to the total proton-proton cross section. The ambiguity in the 
analogy of strings with chromoelectric flux tubes results in a factor
of $\nu =2$ uncertainty in the
relation between the string tension $\kappa$ and the  
product of the colour charge $e$ of the quark and the field strength $\cal E$,
$\nu \kappa = e {\cal E}$ with $\nu \in \lbrack 1,2\rbrack$ \cite{Schon90}.
Two more parameters are the ratio of the inelastic range $R'$ to the full
radius $R$ of the string: $R' /R \approx 0.7$ and the collision time
$\tau_c =0.4$ fm/c fitted to the total and elastic
proton-proton cross sections. Furthermore, the
masses,  degeneracies and
lifetimes taken from \cite{PDG}
have been used 
for the discrete resonances below the mass thresholds, and the 
mesonic and baryonic mass thresholds $M_M = 1.0$ GeV and $M_B = 1.5$
GeV have been chosen.

According to the analogy of the hadronic strings with the
chromoelectric flux tubes, the decay constant $\Lambda $ of the string
is determined by the parameters $\kappa$, $R$, and $\nu$ \cite{Schon90}.
In Table \ref{parsets} we list the parameter sets used for the
simulation of
heavy-ion collision events, including also the corresponding total 
proton-proton
cross sections at high energies and the  decay constants and mean
lifetimes $(T_{l})$ of the strings. The mean lifetimes are determined
for the so-called yo-yo mode \cite{Art83} according to the exponential
decay law, $T_l = \sqrt{ (\ln 2) / \Lambda }$.

\begin{table}
\begin{center}
\begin{tabular}{|c|c|c|c|c||c|c|c|}  \hline
Parameter set & $\kappa$ (GeV/fm) & $R$ (fm) &
  $R'$ (fm) & $\nu$ & $\sigma_{tot}$ (mb) & $\Lambda$ (fm$^{-2}$)
  & $T_{l}$ (fm/c) \\ \hline
(a) & 0.9 & 0.6 & 0.42 & 2.0 & 45 & 1.25 & 0.8 \\ \hline
 (b)& 0.9 & 0.5 & 0.35 & 1.5 & 31 & 0.14 & 2.2 \\ \hline
\end{tabular}
\end{center}
\caption{
\label{parsets}
Parameter sets used for the simulations of heavy-ion collision events.
}
\end{table}

The Dynamical String Model with the parameter sets given in Table
 \ref{parsets} has been tested by simulating elementary hadronization
processes: two-jet events in $e^+ e^- \rightarrow hadrons$ at c.m.
 energies 20 - 50 GeV  \cite{Schon90,hip} and hadronization in
proton-proton collisions at 29 and 200 GeV bombarding energies
\cite{ivanyi1,ivanyi2}.
 
The parameter set (a) is consistent with the total proton-proton cross
sections. It provides a decay constant for which  the simulated results
on the Bose-Einstein correlation of
like-sign pions and on the average charged particle multiplicity
are in good agreement with experimental
data  for 
$e^+ e^- \rightarrow hadrons$ \cite{hip,TPC}. 
Simulated results for the single-particle distributions for the same
process \cite{hip} and for the proton-proton collisions
\cite{ivanyi1,ivanyi2}  are also in
good qualitative agreement with the corresponding data, but the
average charged particle multiplicity in proton-proton collisions is
overestimated nearly by  40\%.

The parameter set (b) has been found  the optimal one in
the simulation for reproducing the single-particle data on
  $e^+ e^- \rightarrow hadrons$ \cite{Schon90}, but it leads to an
  unrealistically small value of the string decay constant and 
  practically no Bose-Einstein correlation of like-sign pions occurs
 in the simulation \cite{hip} using this set. Furthermore, 
the total proton-proton
 cross  section  for high energies is  underestimated by the parameter set (b)
as seen in Table \ref{parsets}. Single-particle data on proton-proton
collisions can be described with a quality similar to that
of the corresponding results for parameter set (a), with a similar
overestimate of the average charged particle multiplicity. It should
be mentioned that
according to the calculations in \cite{Schon90} the parameters of the set
(a) are although not optimal but still in the range which is acceptable
 for describing the single-particle distributions in 
 $e^+ e^- \rightarrow hadrons$.
Thus, the parameter set (a) is preferred on the base of 
comparing
the simulated results with experimental data on the elementary
hadronization processes considered above.

The mean lifetime 0.8 fm/c of  strings  for parameter set (a) is close
 to the value $1.2 \pm 0.1$ fm/c  determined in
 \cite{Bus90}. On the other hand the string radius $R =0.6$ fm
of parameter set (a) is also consistent with the range of its value 
$(0.5 \pm0.1)$ fm, what was established on the base 
of the experimentally observed
strangeness fraction in proton-proton collision \cite{kornel1}.

\section{Simulation of Ultrarelativistic Heavy-Ion Collisions}

The Dynamical String Model described above has been applied to the
simulation of ultrarelativistic heavy-ion collision events. For both
parameter sets 250 collision events were simulated. Hypothetical
 nuclei of mass
number $A=35$ shooted on one another in the c.m.s. were constructed in
 the following
way. The centers of the nucleonic strings (rotating rods of length 0.7
fm)  were positioned at the nodes and the centers of the  cells of a cubical
$3^3$ lattice with lattice spacing 2.1 fm drawn in a sphere of the
nuclear  radius $R_A = r_0 A^{1/3} =3.6$ fm ($r_0 = 1.1$ fm). The
Fermi  motion of the  nucleons has been neglected, their orientations
were chosen randomly according to a uniform distribution in the entire
solid angle. Two such `cubes' with parallel edges were boosted to the
appropriate c.m.s. momenta.  
Central collisions with impact parameters less than 2 fm were
considered. The center of one of the nuclei was chosen according to
a uniform distribution in the  transverse plane,
within a circle of radius of 2 fm around the projection of the center of 
the other nucleus to that plane.
Constructing the hypothetical nuclei 
 in a cubic configuration gives an extra periodic structure
of the nucleus instead of the fluid-like random one. On the other
hand the possible effect of this periodicity is completely neutralised by
choosing different impact parameters for the individual collision
events randomly.  In the numerical simulation no side-effects originating from
 the periodic configuration were seen.

The simulations were performed with the same time steps of $\Delta t =
0.02$ fm/c
as used to fit the total and inelastic radii of the strings and to
perform the test simulations. It is
rather important to take into account the elastic string-string
collisions, since the secondary collisions in ultrarelativistic
heavy-ion collisions play a distinguished role.
 
The simulations were performed with both parameter sets (a) and (b).
The simulated results were transformed back to the laboratory system
and compared with experimental data on   $^{32}$S + $^{32}$S
central collisions at the bombarding energy of $200$ GeV/n in the NA$35$
experiment at CERN \cite{harris,rohrich}. In order to take into
account the
 difference between the mass numbers of
the nuclei in the simulation and the experiment, the simulated distributions
were systematically renormalised by the factor $(32/35)^2$.

\begin{table}
\begin{center}
\begin{tabular}{|c|c|c|c|}  \hline
 &System (lab.energy/n) &  $<{\rm h}^->$ &
 $<{\rm h}^->/B$
\\ \hline
Experiment & S + S ($200$ GeV/n) & $95 \pm 5$ & $1.8 \pm 0.2$ \\ \hline
Experiment & S + Ag ($200$ GeV/n) & $160 \pm 8$ & $1.8 \pm 0.2$ \\ \hline
 Simulation  (a) &
$^{35}$N + $^{35}$N ($200$ GeV/n) & $126 \pm 9$ & $2.4 \pm 0.3$ \\ \hline
 Simulation  (b) & $^{35}$N + $^{35}$N ($200$ GeV/n) &
 $128 \pm 11$ & $2.4 \pm 0.3$ \\ \hline
\end{tabular}
\end{center}
\caption{
\label{multiab}
 Average multiplicities $<{\rm h}^->$ 
of negatively charged hadrons  and the average
multiplicities per total baryon number $B$ for various
 colliding systems.  The simulated  data are
presented for both parameter sets (a) and (b). Here the same number of
net baryons ($B$) is assumed what was measured in the experiment 
S+S.
The experimental
data  are taken  from  \cite{rohrich}.
}
\end{table}

The average multiplicities of the produced hadrons with negative
charge  (supposed to be $\pi^-$ in the experiment) are compared to
 the simulated results in Table
\ref{multiab}.
As the Dynamical String Model does not account for electric charges,
one third of the produced mesons is assumed to be negative based on isospin
arguments.  The simulated average  multiplicity of negatively charged particles
is overestimated by about 30\% as  compared  to the multiplicity
in the reaction $^{32}{\rm S}+^{32}{\rm S}$. This can also be seen from the
data for the averaged negative charged particle  multiplicity per
 participating baryons which varies slightly for different mass
 numbers  \cite{rohrich}. The overestimate  can be the consequence
of overestimating the multiplicities in individual string-string
collisions,  as test simulations for proton-proton collisions have
shown.  Also the complete neglection of the strangeness channel and
the  rather crude treatment of elastic string-string collisions can
affect the average charged particle multiplicity.

\begin{figure}
\epsfxsize=0.9\columnwidth
\epsffile{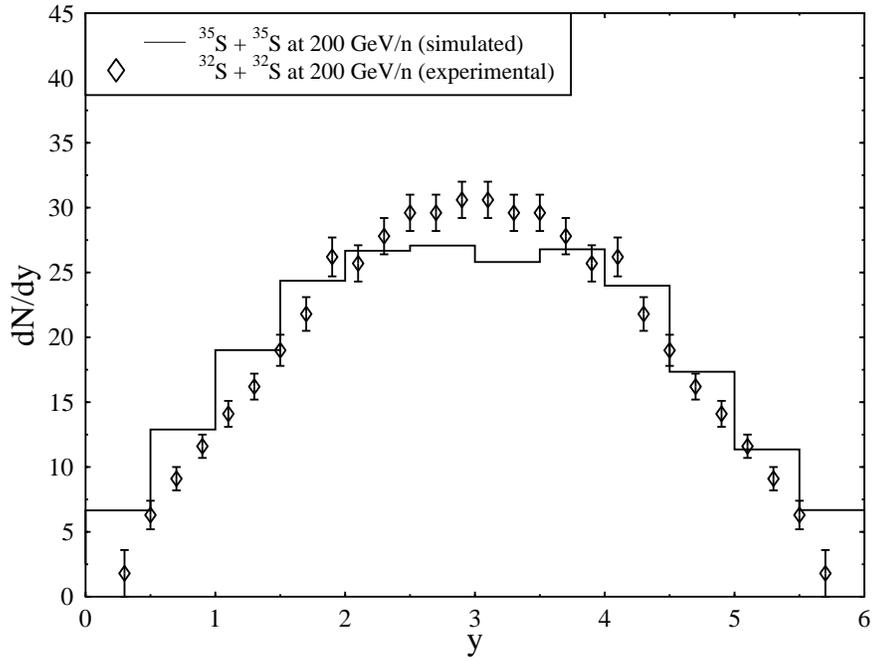}
\epsfxsize=0.9\columnwidth
\epsffile{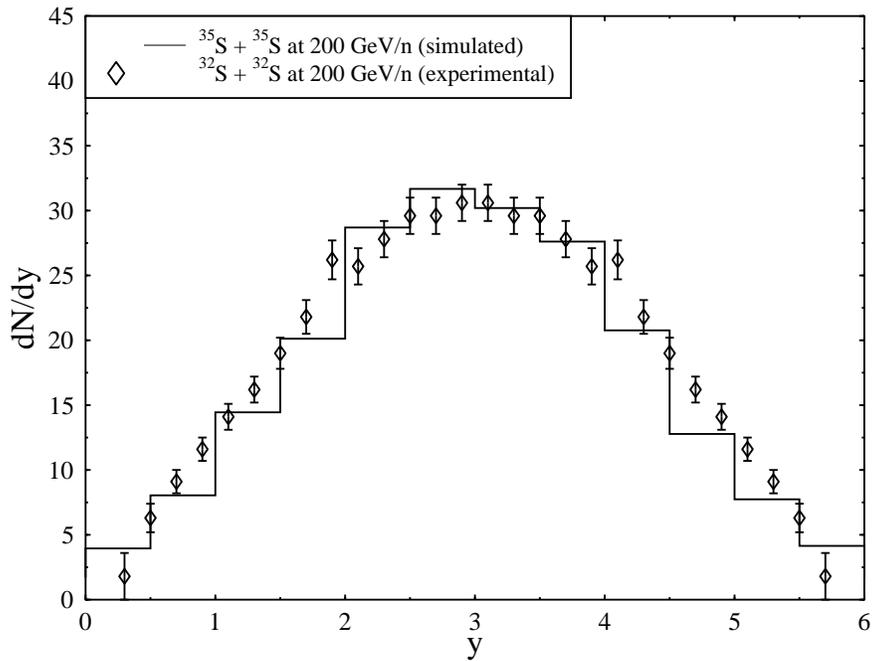}
\caption{\label{rapab}
Rapidity distributions of negatively charged hadrons for the parameter sets
(a) (above) and (b) (below). The experimental data
are from Ref. \cite{harris} which were measured up to mid-rapidity ($y=3$)
and which are simply reflected to higher rapidity supposing symmetry.
}
\end{figure}

\begin{figure}
\epsfxsize=0.9\columnwidth
\epsffile{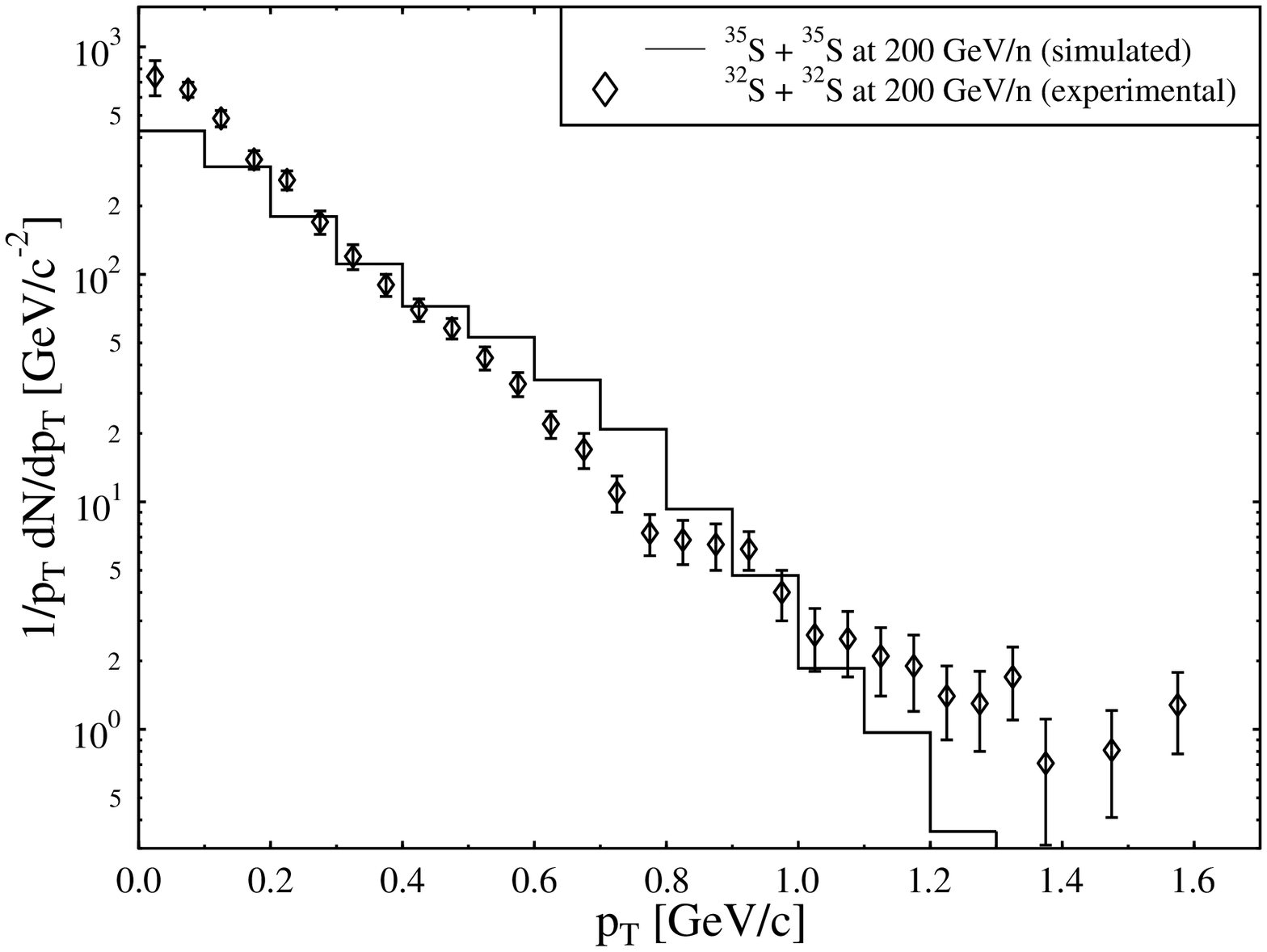}
\epsfxsize=0.9\columnwidth
\epsffile{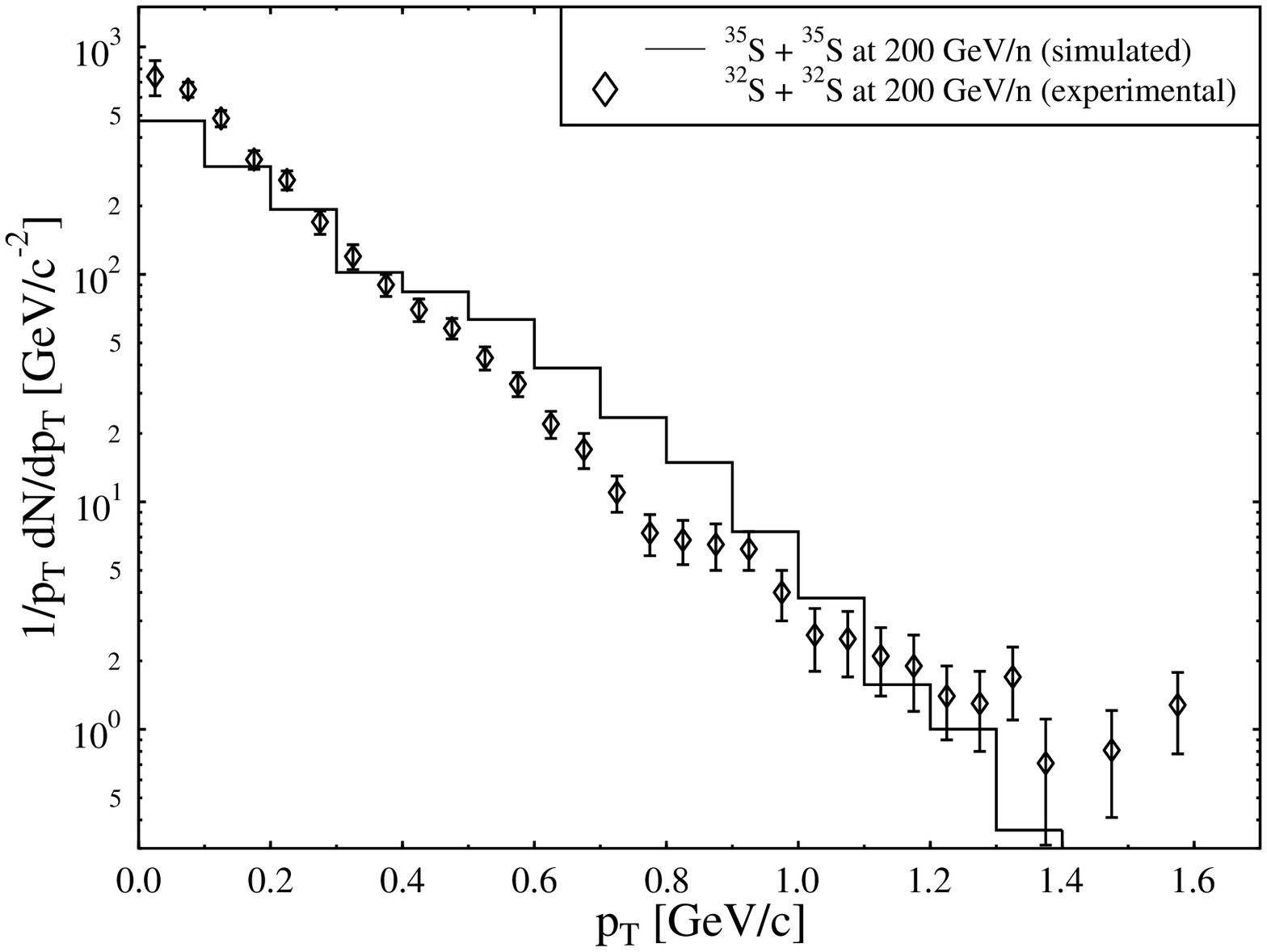}
\caption{\label{trab}
Transverse momentum distributions of negatively charged hadrons for the
parameter
sets (a) (above) and (b) (below). The experimental data are
from Ref. \cite{harris}. The particles are selected
from the central rapidity region ($2<y<3$).
}
\end{figure}

The rapidity and the transverse momentum distributions of  negatively
charged hadrons are shown in
Fig.\ \ref{rapab} and Fig.\ \ref{trab}, respectively. It can be seen that
the experimental spectra are reproduced by the model for both
parameter sets  qualitatively.

\section{Conclusions}

The Dynamical String Model has been generalised in order to simulate
ultrarelativistic heavy-ion collisions at current collider energies.
The initialisation of the incoming nuclei and the discrimination of the
elastic and inelastic scattering of strings are now included.
An effective optimisation of the collision algorithm has been performed
in the numerical code.
In this way, the model is able to simulate nucleus-nucleus collisions at beam
energies of a few hundreds of GeV/nucleon for nuclei with mass numbers up
to around $40$. The simulated results for the reaction $^{32}$S$+^{32}$S at
a beam energy of $200$ GeV/n are in good qualitative agreement with the
experimental data. Therefore, the Dynamical String Model has a predictive power
for ultrarelativistic heavy-ion collisions.

\section*{Acknowledgement}
This work was supported by the Debrecen Research Group in Physics of
the Hungarian Academy of Sciences, by OTKA Project T-023844, by DFG Project
436/UNG/113/123/0,  by GSI and BMBF. The authors are grateful to W.\
Greiner for his kind hospitality at the Institute for Theoretical
Physics, Johann Wolfgang Goethe University. 
K. Sailer thanks for the support of the Alexander von
Humboldt Foundation. B. Iv\'anyi wishes to express his thanks to the DAAD
for their support.

\end{document}